\documentclass[aps,pra,reprint,superscriptaddress]{revtex4-1}
\usepackage{amssymb}
\usepackage{amsfonts}
\usepackage{float}
\usepackage{graphicx}
\usepackage{dcolumn}
\usepackage{bm}
\usepackage{epsfig}
\usepackage{amsmath}
\usepackage{subfigure}
\begin{document}

\title{Tunable optomechanically induced sideband comb}
\author{Jun-Hao Liu}
\address{
Guangdong Provincial Key Laboratory of Nanophotonic Functional Materials and Devices \\
	(School of Information and Optoelectronic Science and Engineering), 
	South China Normal University, Guangzhou 510006, China}
\address{Guangdong Provincial Key Laboratory of Quantum Engineering and Quantum Materials, South China Normal University, Guangzhou 510006, China }
\author{Guangqiang He}
\address{
State Key Laboratory of Advanced Optical Communication Systems and Networks,
Department of Electronic Engineering, Shanghai Jiao Tong University, Shanghai 200240, China}
\author{Qin Wu}
\address{School of Biomedical Engineering, Guangdong Medical University, Dongguan 523808, China }
\author{Ya-Fei Yu}
\address{Guangdong Provincial Key Laboratory of Quantum Engineering and Quantum Materials, South China Normal University, Guangzhou 510006, China }
\author{Jin-Dong Wang}
\email{wangjindong@m.scnu.edu.cn}
\address{Guangdong Provincial Key Laboratory of Quantum Engineering and Quantum Materials, South China Normal University, Guangzhou 510006, China }
\author{Zhi-Ming Zhang}
\email{zhangzhiming@m.scnu.edu.cn}
\address{
	Guangdong Provincial Key Laboratory of Nanophotonic Functional Materials and Devices \\
	(School of Information and Optoelectronic Science and Engineering), 
	South China Normal University, Guangzhou 510006, China}
\begin{abstract}
Cavity optomechanical system can exhibit higher-order sideband comb effect when it is driven by a control field $\omega_{c}$ and a probe field $\omega_{p}$, and works in the non-perturbative regime, as was shown in a previous work [Xiong et al., Opt. Lett. 38, 353 (2013)]. The repetition frequency of such a comb is equal to the mechanical frequency $\omega_{b}$ and is untunable, which limits the precision of the comb. Here we address this problem by driving the system with an additional strong probe field  $\omega_{f}$, and the detuning between $\omega_{f}$ and $\omega_{c}$  is equal to $\omega_{b}/n$ (here $n$ is an integer),  i.e.,  this detuning is a fraction of the mechanical frequency. In this case, we obtain some interesting results. We find that not only the integer-order (higher-order) sidebands, but also the fraction-order sidebands, and the sum and difference sidebands between the integer- and fraction-order sidebands, will appear in the output spectrum. The generated nonlinear sidebands constitute an optomechanically induced sideband comb (OMISC). The frequency range and the repetition frequency of the OMISC are proportional to the sideband cutoff-order number and the sideband interval, respectively. We show that we can extend the frequency range of the OMISC by increasing the intensity of the probe field $\omega_{p}$. More importantly, we can decrease the repetition frequency, and consequently, improve the precision of the OMISC by increasing $n$ and the intensity of the probe field $\omega_{f}$. 
\end{abstract}

\maketitle

\section{Introduction}
Optical frequency comb (OFC) is known as the most accurate ``ruler'' on the earth \cite{1}. Ever since proposed by Chebotaye \cite{2} and  H\"ansch \cite{3}  in 1970s, it has attracted a lot of attention in many study and application areas, such as tests of fundamental physics with atomic clocks \cite{4}, calibration of atomic spectrographs \cite{5}, precision time and frequency transfer \cite{6}, coherent optical communication \cite{7}, molecular fingerprinting \cite{8}, precision ranging \cite{9}, and so on. The original OFC source is a phase-stabilized mode-locked laser \cite{10}. In the time domain the output field of such a mode-locked laser is present as a periodic train of ultrashort pulses, while in the frequency domain this pulse train can be transformed as a Fourier series of equidistant optical frequencies. 

To support more and more broad application space, OFCs have experienced rapid changes over the past 20 years. A lot of new physical systems, such as fiber-based systems \cite{11,12,13}, semiconductor laser platforms \cite{14,15}, microresonator systems \cite{16,17,18,19}, have been studied for achieving the tunable, miniaturized, and  low-powered OFCs. More recently, efforts have also been made in cavity optomechanical system (COMS) \cite{20,21,22}. An interesting effect named higher-order sideband generation (HSG) is discussed by Xiong et al., and they demonstrate that a higher-order sideband comb can be generated in a ultrastrong driven COMS beyond the perturbative approximation \cite{23,24}.

If we consider an OFC as a ``ruler'', the frequency range of the comb will be the length of the ``ruler'', and the repetition frequency will be the minimum scale. For an optomechanically induced sideband comb (OMISC), its frequency range is proportional to the sideband cutoff-order number and its repetition frequency is equal to the sideband interval. In previous works, people mainly focus on increasing the sideband cutoff-order number, for example, ones find that  many physical effects and systems, such as Coulomb effect \cite{25}, atomic ensemble \cite{26}, Casimir effect \cite{27}, and parity-time symmetry structure \cite{28}, can effectively increase the cutoff-order number of the higher-order sideband output from COMS. 

Unlike previous works, we are more interested in tuning or decreasing the sideband interval. Actually in the output spectrum of COMS, the sideband interval is equal to the eigenfrequency of mechanical resonator and is untunable, and we can decrease the sideband interval by decreasing the mechanical frequency. However, owing to the limitation of the processing technology of optomechanical microcavity, it is hard to obtain a microcavity with a small mechanical frequency and a large optomechanical coupling strength. Therefore, ones must explore other ways to decrease the sideband interval. 

In this paper, we address this problem by adding another probe field $\omega_{f}$ into the proposal of Xiong et al. \cite{23}. In particular, we assume that the detuning between $\omega_{f}$ and $\omega_{c}$ is equal to a fraction of the mechanical frequency $\omega_{b}$, i.e, $\omega_{f}-\omega_{c}=\omega_{b}/n$. We find that not only the integer-order (higher-order) sidebands, but also the fraction-order sidebands, and the sum and difference sidebands between integer- and fraction-order sidebands, will appear in the output spectrum. In this case, the sideband interval is equal to $\omega_{b}/n$. So by increasing $n$, we can decrease the sideband interval, and consequently, decrease the repetition frequency and improve the precision of the OMISC.

\begin{figure}[t]
	\centering\includegraphics[width=8cm,height=10cm]{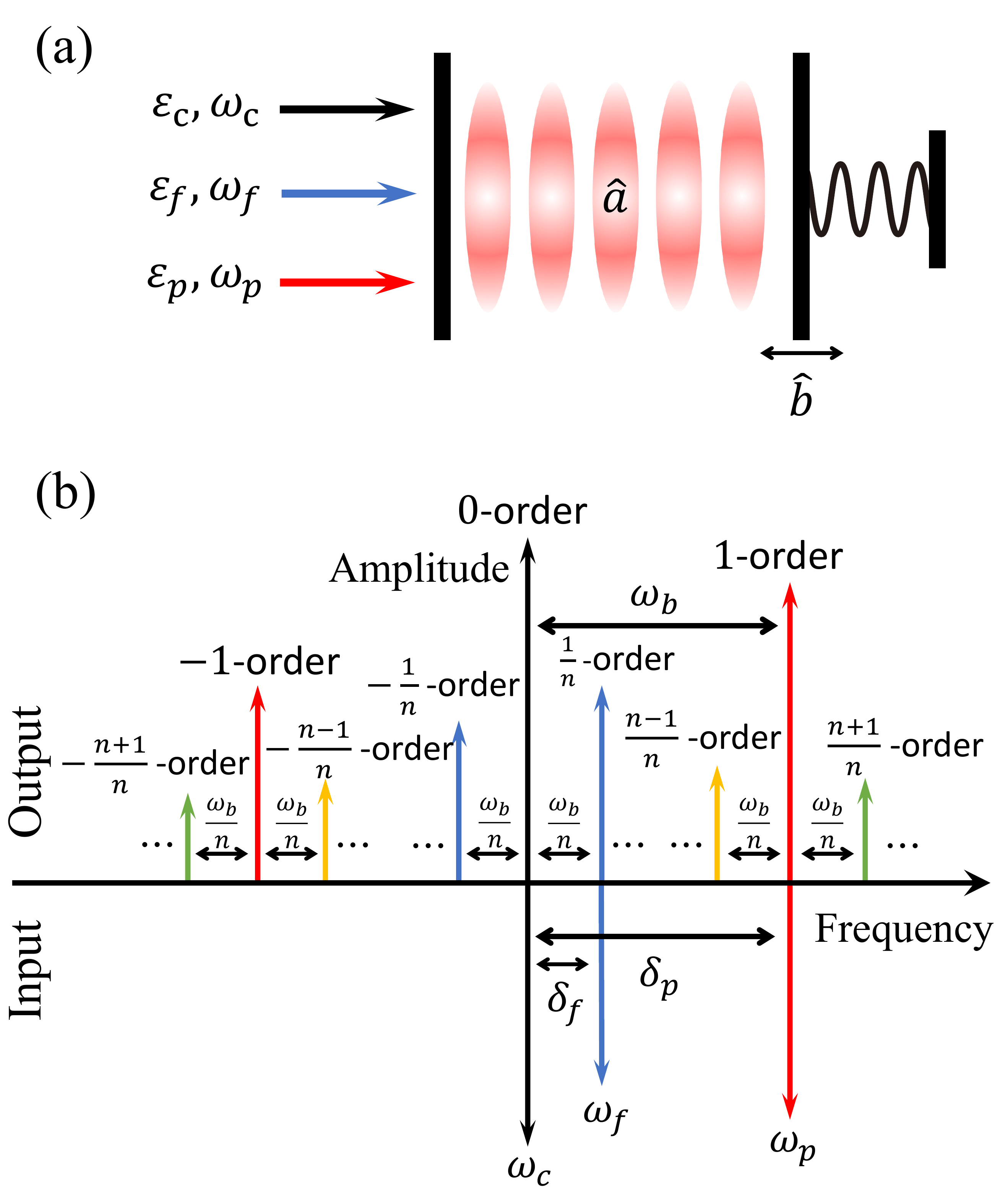}
	\caption{(color online) (a) Standard cavity optomechanical system driven by a control field ($\omega_{c}$) and two probe fields ($\omega_{p}$, $\omega_{f}$). (b) Schematic diagram of the nonlinear sideband generation. We exhibit the integer-order sidebands ($0$-order, $\pm1$-order), the fraction-order sidebands ($\pm\frac{1}{n}$-order), the sum and difference sidebands ($\pm\frac{n\pm1}{n}$-order).}
\end{figure}

 \section{Model and Hamiltonian}
 Our proposed scheme is shown in Fig. 1(a), we consider a standard model of COMS where an optical cavity mode $\hat{a}$ parametrically couples with a mechanical mode $\hat{b}$ via radiation
 pressure.  The cavity mode is pumped by a control field $\omega_{c}$ and two probe fields $\omega_{p}$ and $\omega_{f}$. The Hamiltonian of the system denotes
 \begin{align}
 H=&\hbar \omega_{a}\hat{a}^{\dag }\hat{a}+\hbar \omega _{b}\hat{b}^{\dag }\hat{b} + \hbar g\hat{a}^{\dag }\hat{a}(\hat{b}^{\dag }+\hat{b}) \nonumber \\
 &+i\hbar[\hat{a}^{\dagger}(\varepsilon_{c}e^{-i\omega_{c}t}+\varepsilon_{p}e^{-i\omega_{p}t}+\varepsilon_{f}e^{-i\omega_{f}t})-H.c.],
 \end{align}
 where $\hat{a}$ ($\hat{a}^{\dagger}$) and $\hat{b}$ ($\hat{b}^{\dagger}$) are the bosonic  annihilation (creation) operators for the cavity mode and the mechanical mode with frequencies $\omega _{a}$ and $\omega _{b}$, respectively.  $g$ $=$ $x_{zpf} \omega_{a}/L$ is the single-photon optomechanical coupling strength where $x_{zpf}=\sqrt{\hbar/2M\omega_{b}}$ is the zero-point fluctuation, $M$  is the mass of the mechanical oscillator and $L$ is the length of the cavity. The amplitudes of the driving fields being $\varepsilon _{y}=\sqrt{2\kappa P_{y}/\hbar \omega _{y}}$  ($y=c,p,f$), in which $\kappa$ denotes the cavity damping rate and $P_{y}$ refers to the corresponding input power. 
 
 Based on the Hamiltonian (1), the evolution of the system can be described by the Heisenberg-Langevin equations (in a frame rotating at the frequency $\omega _{c}$):
 \begin{align}
 \frac{d\hat{a}}{dt}=&-(i\Delta _{a}+\kappa )\hat{a}-ig\hat{a}(\hat{b}^{\dag }+\hat{b})  \nonumber\\
 &+\varepsilon _{c}+\varepsilon_{p}e^{-i\delta_{p}t}+\varepsilon_{f}e^{-i\delta_{f}t}+\sqrt{2\kappa }\hat{a}_{in},  \\
 \frac{d\hat{b}}{dt}=&-(i\omega_{b}+\gamma)\hat{b}-ig\hat{a}^{\dagger}\hat{a}+\sqrt{2\gamma }\hat{b}_{in}, 
 \end{align}

 where $\Delta _{a}$ $=$ $\omega _{a}-\omega _{c}$ is the frequency detuning
 between the control field  $\omega _{c}$ and the cavity field $\omega _{a}$.  $\delta_{p}$ $=$ $\omega _{p}-\omega _{c}$ ($\delta_{f}$ $=$ $\omega _{f}-\omega _{c}$) is the frequency detuning between the control field
 $\omega _{c}$  and the probe field $\omega _{p}$ ($\omega _{f}$). $\gamma$ is the mechanical damping rate. $\hat{a}_{in}$ and $\hat{b}_{in}$ are respectively the fluctuation operations corresponding to the cavity mode and the mechanical mode with zero mean value, i.e., $\langle \hat{a}_{in}\rangle$ $=$ $\langle \hat{b}_{in}\rangle$ $=$ $0$.  We focus on the mean response of the system to the probe fields, thus in the following we turn to calculate the evolutions of the expectation values of the system operators $\hat{a}$ and $\hat{b}$, and we denote $\left\langle 
 \hat{a}\right\rangle \equiv \alpha$, $\langle \hat{b}\rangle \equiv \beta$. By using the mean-field approximation, i.e., $\langle \hat{a}\hat{b}\rangle =\left\langle 
 \hat{a}\right\rangle \langle \hat{b}\rangle$, the dynamical equations of the system can be derived from Eqs.(2)-(3) as
 \begin{align}
 \frac{d\alpha}{dt}=&-(i\Delta _{a}+\kappa )\alpha-ig\alpha(\beta+\beta^{*}) \nonumber\\
 &+\varepsilon _{c}+\varepsilon_{p}e^{-i\delta_{p}t}+\varepsilon_{f}e^{-i\delta_{f}t}, \\
 \frac{d\beta}{dt}=&-(i\omega_{b}+\gamma)\beta-ig\left|\alpha \right|^{2}.
 \end{align}
 We recall that the solutions of the above nonlinear equations have been addressed in previous works for the following three parametric conditions: (i) $\varepsilon_{f}=0, \varepsilon_{p}\neq0, \varepsilon_{p}/\varepsilon_{c}\ll1$, under this circumstance the equations can be solved by using the so-called linearization method, and the Stokes and anti-Stokes processes are discussed \cite{29}. (ii) $\varepsilon_{f}\neq0, \varepsilon_{p}\neq0, \varepsilon_{p}/\varepsilon_{c}\ll1, \varepsilon_{f}/\varepsilon_{c}\ll1$, in this case the linearization is practicable, and the sum and difference sidebands effects are investigated \cite{30,31}.  (iii) $\varepsilon_{f}=0, \varepsilon_{p}\neq0, \varepsilon_{p}\sim\varepsilon_{c}$, in this regime the linearization is inapplicable, and the higher-order sideband effect is studied by using the numerical calculation \cite{23}. 

\section{Methods}
In this paper, we study the fourth circumstance, i.e., $\varepsilon_{p}\neq0, \varepsilon_{f}\neq0, \varepsilon_{f}\sim\varepsilon_{p}\sim\varepsilon_{c}$, in this regime the expectation value $x$ ($x=\alpha,\beta$) can be written as
\begin{eqnarray}
&&x=x_{0}+x_{p}+ x_{f}+x_{s}+ x_{d},
\end{eqnarray}
where  $x_{0}$ denotes the steady-state solution when $\varepsilon_{p}=\varepsilon_{f}=0$. The optomechanical nonlinearity [corresponding to the terms $-ig\left|\alpha \right|^{2}$ and $-ig\alpha(\beta^{* }+\beta)$] will result in the generation of new photons with different frequencies. When the laser-fields are incident upon the cavity, the driving energy will be transferred to a series of sidebands with new frequencies, which can be expressed as
\begin{eqnarray}
x_{p}=\sum_{j=1}^{N}x_{p+}^{(j)}e^{-ij\delta_{p}t}+x_{p-}^{(j)}e^{ij\delta_{p}t},\\
x_{f}=\sum_{k=1}^{M}x_{f+}^{(k)}e^{-ik\delta_{f}t}+x_{f-}^{(k)}e^{ik\delta_{f}t},
\end{eqnarray}
similarly to the sum and difference frequency generations in the nonlinear medium, a series of  sum and difference sidebands will be generated, which can be written as 
\begin{eqnarray}
x_{s}=\sum_{j=1}^{N}\sum_{k=1}^{M}x_{s+}^{(j,k)}e^{-i(j\delta_{p}+k\delta_{f})t}+x_{s-}^{(j,k)}e^{i(j\delta_{p}+k\delta_{f})t}, \\ x_{d}=\sum_{j=1}^{N}\sum_{k=1}^{M}x_{d+}^{(j,k)}e^{-i(j\delta_{p}-k\delta_{f})t}+x_{d-}^{(j,k)}e^{i(j\delta_{p}-k\delta_{f})t}.
\end{eqnarray}

By solving Eqs.(4)-(5) with the help of the ansatz (6), and using the input-output relation \cite{32}  $S_{out}=S_{in}-\sqrt{2\kappa}\alpha$ ($S_{in}=\varepsilon _{c}+\varepsilon_{p}e^{-i\delta_{p}t}+\varepsilon_{f}e^{-i\delta_{f}t}$), we can finally obtain the output spectrum of the system
\begin{eqnarray}
S_{out}=S_{0}+S_{p}+ S_{f}+S_{s}+S_{d},
\end{eqnarray}
in which we have assumed that the detunings between the probe fields and the control field satisfy: (i) $\delta_{p}=\omega_{b}$;
(ii) $\delta_{f}=\frac{\omega_{b}}{n}$ ($n$ is a positive integer number). It should be noted that the spectrum obtained have shifted the frequency $\omega_{c}$, because Eqs.(4)-(5) describe the evolution of the optical field in a frame rotating at the control frequency. The output spectrum includes the following five parts, as shown in Fig.1(b), the first part $S_{0}=\varepsilon_{c}-\sqrt{2\kappa}\alpha_{0}$ is  the $0$-order sideband which corresponds to the control field $\omega_{c}$. The second and third parts denote the non-zero integer- and fraction-order sidebands, respectively, which can be expressed as
\begin{eqnarray}
S_{p}=\sum_{j=1}^{N}A_{p+}^{(j)}e^{-ij\omega_{b}t}+A_{p-}^{(j)}e^{ij\omega_{b}t},\\ 
S_{f}=\sum_{k=1}^{M}A_{f+}^{(k)}e^{-i\frac{k}{n}\omega_{b}t}+A_{f-}^{(k)}e^{i\frac{k}{n}\omega_{b}t},
\end{eqnarray} 
we will see a series of sidebands appear at $\omega=\pm j\omega_{b}$ and $\omega=\pm \frac{k}{n}\omega_{b}$, in which $A_{p+}^{(1)}$ and $A_{f+}^{(1)}$ is called the $1$- and $\frac{1}{n}$-order sideband, which  correspond to the probe field $\omega_{p}$ and $\omega_{f}$, respectively. The fourth and fifth parts describe the sum and difference sidebands, respectively, which are given by
\begin{eqnarray}
S_{s}=\sum_{j=1}^{N}\sum_{k=1}^{M}A_{s+}^{(j,k)}e^{-i\frac{jn+k}{n}\omega_{b}t}+A_{s-}^{(j,k)}e^{i\frac{jn+k}{n}\omega_{b}t}, \\ S_{d}=\sum_{j=1}^{N}\sum_{k=1}^{M}A_{d+}^{(j,k)}e^{-i\frac{jn-k}{n}\omega_{b}t}+A_{d-}^{(j,k)}e^{i\frac{jn-k}{n}\omega_{b}t}.
\end{eqnarray} 
That means a series of new sidebands will appear at $\omega=\pm \frac{jn\pm k}{n}\omega_{b}$, and we call $A_{s\pm}^{(j,k)}$ and $A_{d\pm}^{(j,k)}$ as the $\pm \frac{jn+ k}{n}$- and $\pm \frac{jn- k}{n}$-order sidebands, respectively.  If all the generated nonlinear sidebands appear in the output spectrum with equal interval, then they will form an OFC, and we call it as optomechanically induced sideband comb (OMISC). 

It can be seen that it is very difficult and tedious to give the analytical solutions of the sidebands for every orders. To verify our theory and exhibit all the generated sidebands, a more convenient practice is to use the numerical calculation, and we adopt the Runge-Kutta method to solve Eqs.(4)-(5), then the output spectrum can be obtained by using the fast Fourier transform. The used parameters are chosen based on the recent experiment \cite{33} and can be achieved under current technology: $\omega _{b}/2\pi=51.8$ MHz and $\gamma /2\pi=41$ kHz (quality factor $Q_{m}$ $=$ $1.26\times 10^{3}$), $m=20$ ng, $\Delta_{a}=\omega_{b}$, $\kappa/2\pi =15$ MHz, $g/2\pi=1$  kHz, the wavelength of the control field $\lambda_{c}=795$ nm, and $\varepsilon_{c}=3\times10^{3}$ GHz.

\begin{figure}[tbp]
	\centering\includegraphics[width=8.5cm,height=6.4cm]{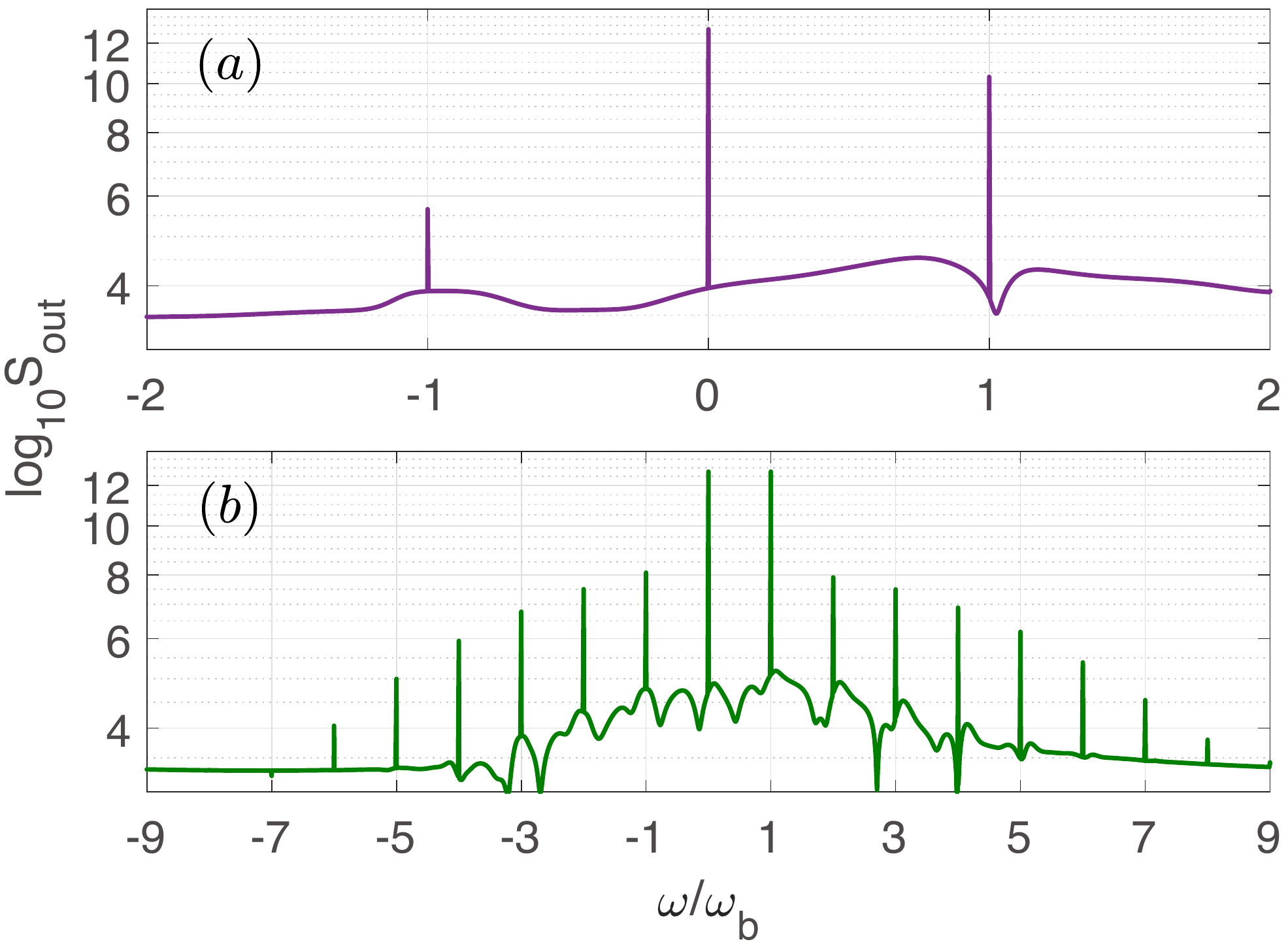}
	\caption{(color online) The integer-order sideband spectra for different $\varepsilon_{p}$, (a) $\varepsilon_{p}=9$ GHz, (b) $\varepsilon_{p}=3\times10^{3}$ GHz, the other parameters are stated in the text.}
\end{figure}

\section{Results and Discussions}
To exhibit the nonlinear sidebands output from COMS and the generation of the OMISC, we firstly assume that $\varepsilon_{f}=0$ and show the integer-order (higher-order) sideband spectra in Fig. 2. We can see that when $\varepsilon_{p}$ is weak ($\varepsilon_{p}/\varepsilon_{c}=3\times10^{-2}$), there are only $0$-order and $\pm1$-order sidebands in the output spectrum. If we increase $\varepsilon_{p}$ and when $\varepsilon_{p}$ is strong ($\varepsilon_{p}/\varepsilon_{c}=1$), the amplitudes of $\pm1$-order sidebands are strengthened, and the higher integer-order sidebands appear, the positive and negative sidebands end up at the orders of +8 and -6, respectively. Furthermore, for higher order of the sideband, the amplitude is smaller. In general, both the sideband cutoff-order number and the amplitude will gradually increase with the increasing of $\varepsilon_{p}$. However, the sideband interval does not change and is always $\omega_{b}$.
From the point of view of OFC, the integer-order sidebands discussed above constitute an OMISC. The frequency range $f_{r}$ can be extended by increasing $\varepsilon_{p}$. For example, with the increase of $\varepsilon_{p}$, $f_{r}$ can be extended from $[-\omega_{b},+\omega_{b}]$ (Fig. 2(a)) to  $[-6\omega_{b},+8\omega_{b}]$ (Fig. 2(b)), whereas the repetition frequency $f_{rep}$ remains $\omega_{b}$. 
\begin{figure}[t]
	\centering\includegraphics[width=8.5cm,height=9cm]{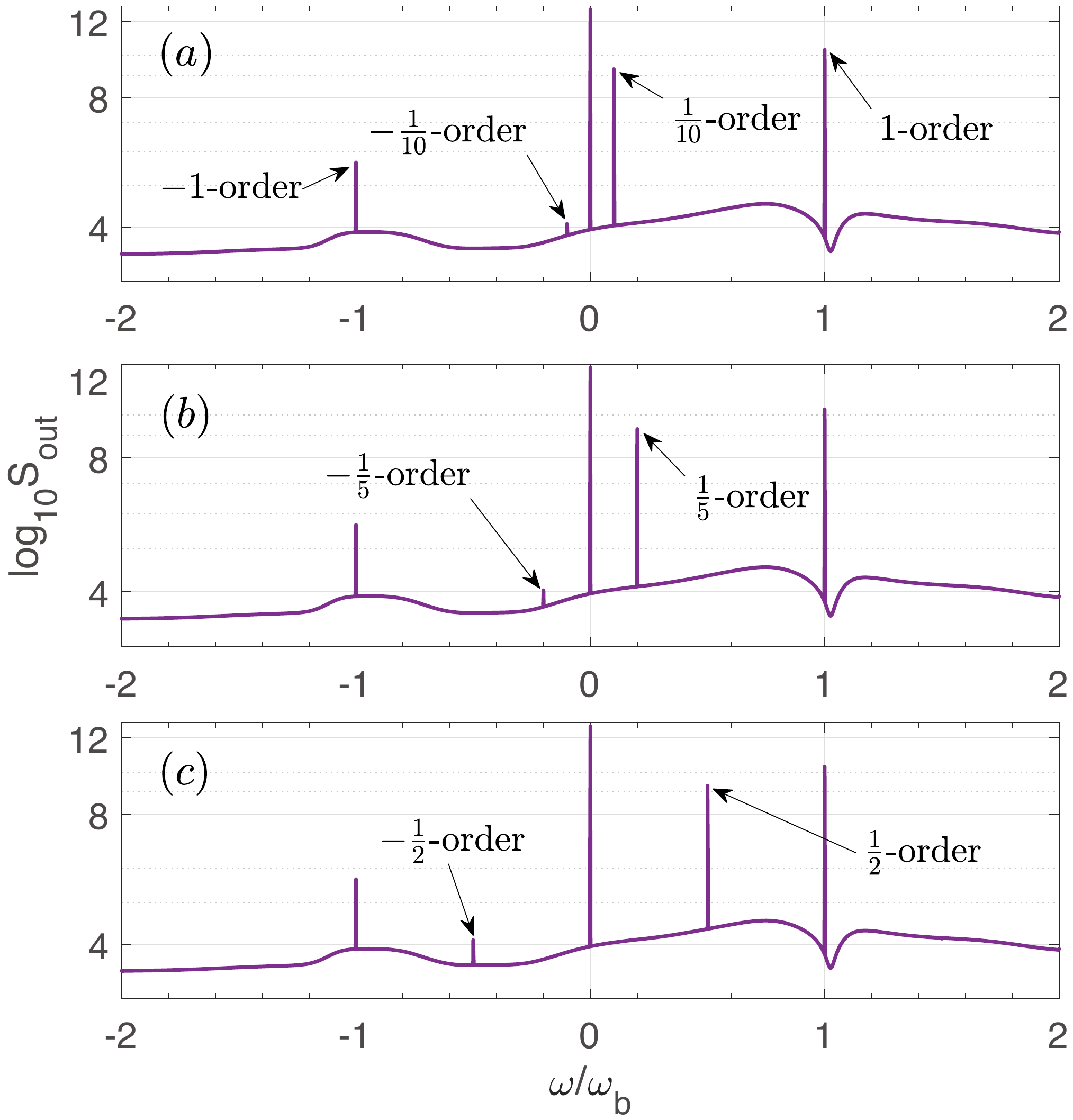}
	\caption{(color online) The integer-order and fraction-order sideband spectra for different $n$, (a) $n=10$, (b) $n=5$, (c) $n=2$. $\varepsilon_{f}=9\times10^{-1}$ GHz, and the other parameters are the same as that in Fig. 2(a).}
\end{figure}

\begin{figure}[t]
	\centering\includegraphics[width=8.5cm,height=9cm]{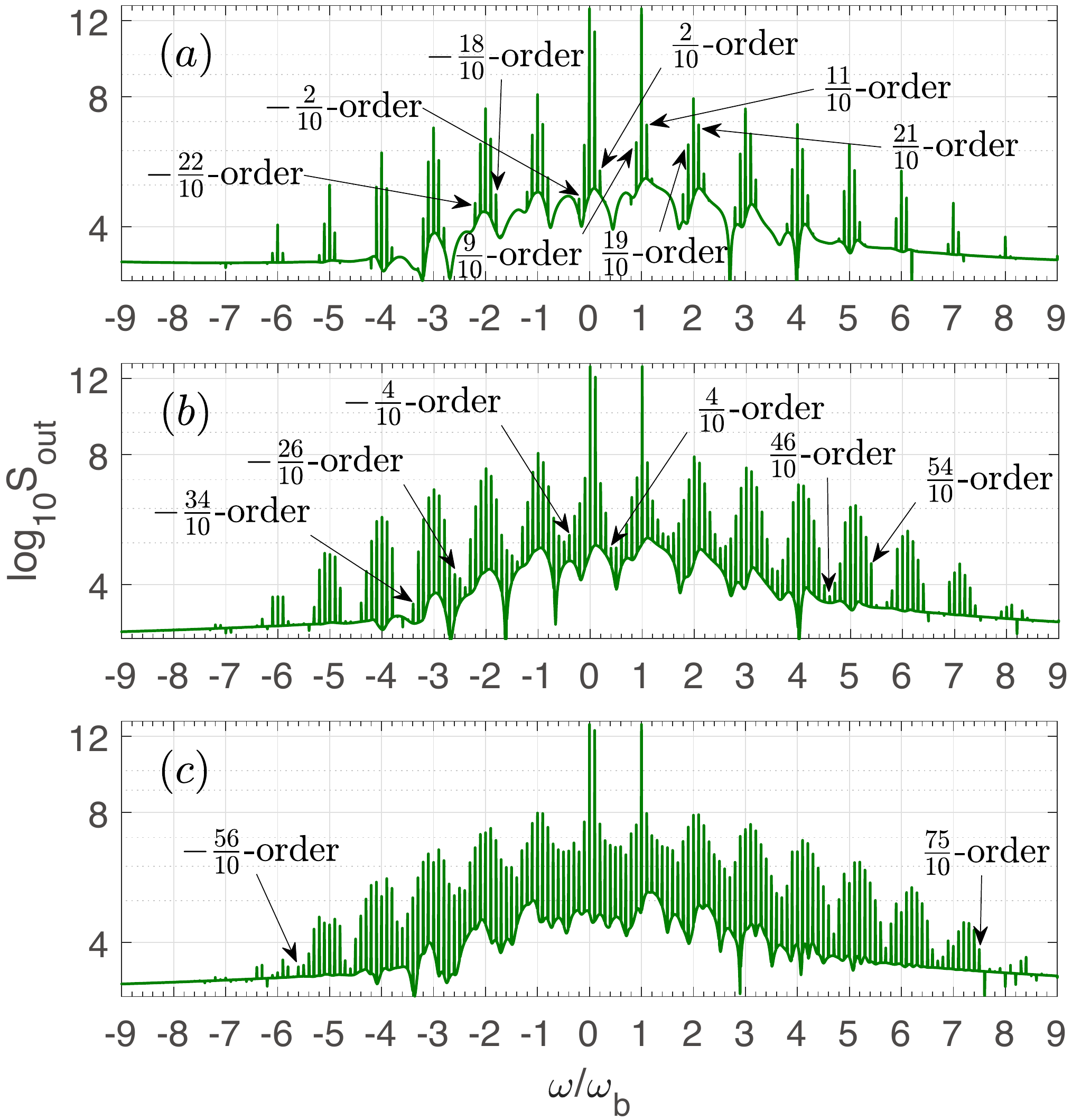}
	\caption{(color online) The higher integer-order, fraction-order, sum, and difference sideband spectra for different $\varepsilon_{f}$, (a) $\varepsilon_{f}=9\times10^{1}$ GHz, (b) $\varepsilon_{f}=6\times10^{2}$ GHz, (c) $\varepsilon_{f}=1.2\times10^{3}$ GHz. $n=10$, and the other parameters are the same as that in Fig. 2(b). }
\end{figure}

Below we show the integer-order and fraction-order sideband spectra for different $n$, as shown in Fig. 3, we consider that another probe field $\omega_{f}$ is inputted into the cavity with a very small amplitude ($\varepsilon_{f}/\varepsilon_{c}=3\times10^{-3}$), and the other conditions are the same as that in Fig. 2(a). It can be seen that the integer-order sidebands have hardly changed, while two new fraction-order sidebands appear, and their positions depend on $n$. Especially when $n=2$, the new generated $\pm\frac{1}{2}$-order sidebands and the pre-existing $0$-order and $\pm1$-order sidebands constitute an OMISC. In this case, the repetition frequency of the OMISC becomes a half of the mechanical frequency, i.e., $f_{rep}=\omega_{b}/2$. However, the frequency range $f_{r}$ of the OMISC remains [$-\omega_{b},\omega_{b}$]. 

Figure 4 displays the higher integer-order, fraction-order, sum and difference sideband spectra for different $\varepsilon_{f}$, and the other conditions are the same as that in Fig. 2(b). When $\varepsilon_{f}/\varepsilon_{c}=3\times10^{-2}$, the amplitudes of $\pm \frac{1}{10}$-order sidebands are obviously enhanced compared with that in Fig. 3(a), and the sum and difference sidebands between $\pm\frac{1}{10}$-order and the integer-order sidebands, such as $\pm\frac{11}{10}$-order, $\pm\frac{9}{10}$-order, $\pm\frac{19}{10}$-order, $\pm\frac{21}{10}$-order sidebands, etc., can be seen in the output spectrum. Furthermore, the higher fraction-order sidebands, such as $\pm \frac{2}{10}$-order sidebands come out, and the sum and difference sidebands between $\pm \frac{2}{10}$-order and the integer-order sidebands can be also seen, for example, $\pm\frac{18}{10}$-order, $-\frac{22}{10}$-order sidebands, and so on. When $\varepsilon_{f}/\varepsilon_{c}=2\times10^{-1}$, there are more higher fraction-order sidebands, such as $\pm\frac{4}{10}$-order sidebands, and more sum and difference sidebands, such as $-\frac{26}{10}$-order, $\frac{46}{10}$-order sidebands. All the generated nonlinear sidebands (from $-\frac{34}{10}$-order to $\frac{54}{10}$-order) constitute an OMISC with the frequency range $f_{r}=[-\frac{34}{10}\omega_{b},+\frac{54}{10}\omega_{b}]$ and repetition frequency $f_{rep}=\omega_{b}/10$. If we continue to increase $\varepsilon_{f}$ and when $\varepsilon_{f}/\varepsilon_{c}=4\times10^{-1}$, we can see that more sum and difference sidebands appear, for example, $-\frac{56}{10}$-order, $\frac{75}{10}$-order sidebands. From the angle of the OMISC, in this case $f_{r}$ is approximately equal to that in Fig. 2(b) while $f_{rep}$ is decreased by one order of magnitude.

\section{Conclusions}
In summary, we have shown the optomechanical induced nonlinear sideband generation in a standard optomechanical system driven by three laser-fields consisting of a control field $\omega_{c}$ and two probe field $\omega_{p}$ and $\omega_{f}$. The detuning between $\omega_{c}$ and $\omega_{p}$ ($\omega_{f}$) is equal to the mechanical frequency $\omega_{b}$ ($\omega_{b}/n$).  The output sidebands can come into being a tunable OMISC. We show that the sidebands cutoff-order number can be extended by increasing the intensity of the probe field $\omega_{p}$, which results in the enlargement of the frequency range of the OMISC. In addition, the sidebands interval can be decreased by increasing $n$ and the intensity of the probe field $\omega_{f}$, which leads to the diminution of the repetition frequency of the OMISC. We conclude that the challenge to achieve the tunable OFC based on COMS will be greatly aided by our work.
\section*{Funding} National Natural Science Foundation of China (61941501, 61775062, 11574092, 61475099, 61378012, and 91121023), Program of State Key Laboratory of Quantum Optics and Quantum Optics Devices (KF201701), Doctoral Program of Guangdong Natural Science Foundation (2018A030310109).


\begin{thebibliography}{99}
	
	\bibitem{1} T. Fortier and E. Baumann,\textquotedblleft 20 years of developments in optical frequency comb technology and applications,\textquotedblright Commun. Phys. \textbf{2}, 1 (2019).

\bibitem{2} Y. V. Baklanov and V. P. Chebotayev, \textquotedblleft Narrow resonances of two-photon absorption of super-narrow pulses in a gas,\textquotedblright Appl. Phys. A \textbf{12}, 97 (1977). 

\bibitem{3} J. N. Eckstein, A. I. Ferguson, and T. W. H\"ansch, \textquotedblleft High-resolution two-photon spectroscopy with picosecond light pulses,\textquotedblright Phys. Rev. Lett. \textbf{40},  847 (1978).

\bibitem{4} T. Rosenband, D. B. Hume, P. O. Schmidt, C. W. Chou, A. Brusch, L. Lorini, et al., \textquotedblleft Frequency ratio of Al$^{+}$ and Hg$^{+}$ single-ion optical clocks; metrology at the 17th decimal place,\textquotedblright Science \textbf{319}, 1808 (2008).

\bibitem{5} M. T. Murphy, T. Udem, R. Holzwarth,  A. Sizmann, L. Pasquini, C. Araujo-Hauck, et al., \textquotedblleft High-precision wavelength calibration of astronomical spectrographs with laser frequency combs,\textquotedblright Mon. Not. R. Astron. Soc. \textbf{380}, 839 (2007).


\bibitem{6} F. R. Giorgetta, W. C. Swann,  L. C. Sinclair, E. Baumann,  I. Coddington, and  N. R. Newbury, \textquotedblleft Optical two-way time and frequency transfer over free space,\textquotedblright Nat. Photon.  \textbf{7}, 434 (2013). 

\bibitem{7}  P. Marin-Palomo,  J. N. Kemal, M. Karpov, A. Kordts, J. Pfeifle, M. H. P. Pfeiffer, et al., \textquotedblleft Microresonator-based solitons for massively parallel coherent optical communications,\textquotedblright Nature, \textbf{546}, 274 (2017). 

\bibitem{8} S. A. Diddams, L. Hollberg,  and V.  Mbele, \textquotedblleft Molecular fingerprinting with the resolved modes of a femtosecond laser frequency comb,\textquotedblright Nature \textbf{445}, 627 (2007).

\bibitem{9} K. Minoshima and H. Matsumoto,  \textquotedblleft High-accuracy measurement of 240-m distance in an optical tunnel by use of a compact femtosecond laser,\textquotedblright Appl. Opt.  \textbf{39},  5512 (2000).

\bibitem{10} T. M. Fortier, P. A. Roos, D. J. Jones,  S. T. Cundiff, R. D. R. Bhat, and J. E. Sipe,  \textquotedblleft Carrier-envelope phase-controlled quantum interference of injected photocurrents in semiconductors,\textquotedblright Phys. Rev. Lett. \textbf{92}, 147403 (2004). 

\bibitem{11} S. Schilt and T. Südmeyer,  \textquotedblleft Carrier-envelope offset stabilized ultrafast diode-pumped solid-state lasers,\textquotedblright Appl. Sci. \textbf{5}, 787 (2015).

\bibitem{12} W. Xia and X. Chen, \textquotedblleft Recent developments in fiber-based optical frequency comb and its applications,\textquotedblright Meas. Sci. Technol. \textbf{27}, 041001 (2016).

\bibitem{13} S. Duval, M. Bernier, V. Fortin, J. Genest, M. Piché,  and R. Vallée,  \textquotedblleft Femtosecond fiber lasers reach the mid-infrared,\textquotedblright Optica  \textbf{2}, 623 (2015). 

\bibitem{14} S. M. Link, A. Klenner, M. Mangold, C. A. Zaugg, M. Golling, B. W. Tilma, and U.  Keller,  \textquotedblleft Dual-comb modelocked laser,\textquotedblright Opt. Express \textbf{23}, 5521 (2015). 

\bibitem{15} L. A. Sterczewski, J. Westberg, Y. Yang, D. Burghoff, J. Reno, Q. Hu, and G. Wysocki, \textquotedblleft Terahertz hyperspectral imaging with dual chip-scale combs,\textquotedblright Optica \textbf{6}, 766 (2019). 

\bibitem{16} P. Del’Haye, A. Schliesser, O. Arcizet, T. Wilken, R. Holzwarth, and T. J. Kippenberg, \textquotedblleft Optical frequency comb generation from a monolithic microresonator,\textquotedblright Nature \textbf{450}, 1214 (2007).

\bibitem{17} M. Kourogi, K. Nakagawa, and M. Ohtsu, \textquotedblleft Wide-span optical frequency comb generator for accurate optical frequency difference measurement,\textquotedblright IEEE J. Quantum Electron. \textbf{29}, 2693 (1993).

\bibitem{18} V. Brasch, E. Lucas, J. D. Jost, M.  Geiselmann, and T. J. Kippenberg, \textquotedblleft Self- referenced photonic chip soliton kerr frequency comb,\textquotedblright  Light-Sci. Appl. \textbf{6}, e16202 (2017).

\bibitem{19}  T. Herr, V. Brasch, J. D. Jost, C. Y. Wang, N. M. Kondratiev, M. L. Gorodetsky, and T. J. Kippenberg, \textquotedblleft Temporal solitons in optical microresonators,\textquotedblright Nat. Photon. \textbf{8}, 145 (2014). 

\bibitem{20}  A. A. Savchenkov, A. B. Matsko, V. S. Ilchenko, I. Solomatine, D. Seidel, and  L. Maleki, \textquotedblleft Tunable optical frequency comb with a crystalline
whispering gallery mode resonator,\textquotedblright Phys. Rev. Lett.  \textbf{101}, 093902 (2008). 

\bibitem{21}  A. A. Savchenkov, A. B. Matsko, V. S. Ilchenko, D. Seidel, and L. Maleki, \textquotedblleft Surface acoustic wave opto-mechanical oscillator and frequency comb generator,\textquotedblright Opt. Lett. \textbf{36},  3338 (2011). 

\bibitem{22}  M. A. Miri, G. D’Aguanno, and A. Alù, \textquotedblleft Optomechanical frequency combs,\textquotedblright  New J. Phys. \textbf{20},  043013 (2018). 

\bibitem{23} H. Xiong, L. G. Si, X. Y. Lü, X. X. Yang, and Y. Wu, \textquotedblleft Carrier-envelope phase-dependent effect of high-order sideband generation in ultrafast driven optomechanical system,\textquotedblright
Opt. Lett. \textbf{38}, 353 (2013).

\bibitem{24} H. Xiong, L. G. Si, X. Y. Lü, X. X. Yang, and Y. Wu, \textquotedblleft Nanosecond-pulse-controlled higher-order sideband comb in a GaAs optomechanical disk resonator in the non-perturbative regime,\textquotedblright Ann. Phys-New York \textbf{349}, 43 (2014).

\bibitem{25} C. Kong, H. Xiong, and Y. Wu, \textquotedblleft Coulomb-interaction-dependent effect of high-order sideband generation in an optomechanical system,\textquotedblright  Phys. Rev. A \textbf{95} , 033820 (2017).

\bibitem{26} Z. X. Liu, H. Xiong, and Y. Wu,  \textquotedblleft Generation and amplification of a high-order sideband induced by two-level atoms in a hybrid optomechanical system,\textquotedblright Phys. Rev. A \textbf{97}, 013801 (2018).

\bibitem{27} J. Yao, Y. F. Yu, and Z. M. Zhang,  \textquotedblleft Effects of Casimir force on high-order sideband generation in an optomechanical system,\textquotedblright Chin. Opt. Lett. \textbf{16}, 111201(2018).

\bibitem{28} L. Y. He, \textquotedblleft Parity-time-symmetry–enhanced sideband generation in an optomechanical system,\textquotedblright Phys. Rev. A \textbf{99}, 033843 (2019).

\bibitem{29} S. Huang and G. S. Agarwal,  \textquotedblleft Normal-mode splitting and antibunching in Stokes and anti-Stokes processes in cavity optomechanics: radiation-pressure-induced four-wave-mixing cavity optomechanics,\textquotedblright Phys. Rev. A \textbf{81}, 033830 (2010).

\bibitem{30} H. Xiong, L. G. Si, X. Y. L\"u,and Y. Wu, \textquotedblleft Optomechanically induced sum sideband generation,\textquotedblright Opt. Express, \textbf{24}, 5773 (2016).

\bibitem{31} H. Xiong, Y. W. Fan, X. X. Yang, and Y. Wu, \textquotedblleft Radiation pressure induced difference-sideband generation beyond linearized description,\textquotedblright Appl. Phys. Lett. \textbf{109}, 061108  (2016).

\bibitem{32} C. W. Gardiner and M. J. Collett, \textquotedblleft Input and output in damped quantum systems: Quantum stochastic differential equations and the master equation,\textquotedblright Phys. Rev. A \textbf{31}, 3761 (1985).

\bibitem{33} S. Weis, R. Rivi\'ere, S. Del\'eglise, E. Gavartin, O. Ar- cizet, A. Schliesser, and T. J. Kippenberg, \textquotedblleft Optomechanically induced transparency,\textquotedblright Science \textbf{330}, 1520 (2010).
\end{thebibliography}
\end{document}